\documentclass[12pt]{iopart}

\usepackage{graphicx}
\usepackage{hyperref}
\usepackage{movie15}
\usepackage[numbers,square,sort&compress]{natbib}

\begin{document}

\title[Ion-neutral interactions in solar plasmas]{On the effects of ion-neutral interactions in solar plasmas}

\author{Elena Khomenko$^{1,2}$}

\address{$^1$Instituto de Astrof\'{\i}sica de Canarias, 38205 La Laguna, Tenerife, Spain \\
$^2$Departamento de Astrof\'{\i}sica, Universidad de La Laguna, 38205, La Laguna, Tenerife, Spain }

\ead{khomenko@iac.es}
\vspace{10pt}
\begin{indented}
\item[]June 2016
\end{indented}

\begin{abstract}
Solar photosphere and chromosphere are composed of weakly ionized plasma for which collisional coupling decreases with height. This implies a breakdown of some hypotheses underlying magnetohydrodynamics at low altitudes and gives rise to non-ideal MHD effects such as ambipolar diffusion, Hall effect, etc. Recently, there has been progress in understanding the role of these effects for the dynamics and energetics of the solar atmosphere. There are evidences that such phenomena as wave propagation and damping, magnetic reconnection, formation of stable magnetic field concentrations, magnetic flux emergence, etc. can be affected. This paper reviews the current state-of-the-art of multi-fluid MHD modeling of the coupled solar atmosphere. 
\end{abstract}

%
%
%
%
%

\section{Introduction}

The classical approach in describing the interaction between the magnetic field and the plasma in the solar atmosphere consists in the application of magnetohydrodynamic equations, using different levels of simplifications \cite{Asplund+etal2000, Khomenko+Collados2006, Cheung+etal2007, Moreno-Insertis+etal2008}. Nevertheless the solar lower atmosphere (photosphere and chromosphere) is composed of a weakly ionized plasma. The low temperatures around the photospheric temperature minimum lead to an ionization fraction around $n_i/n\approx10^{-4}$ at higher altitude (where 1 means fully ionized plasma). In the chromosphere the ionization fraction increases, but always remains below 1. As the atmosphere becomes less dense with height, its charge-neutral collisional coupling decreases and each species composing the solar plasma stop exhibiting the collective behavior. Since different forces act on charged and neutral particles, and also on particles with different mass, the breakdown of collective behavior leads to an appearance of net collisional forces between different species. The differential behavior of neutrals and ions gives rise to charge separation and ambipolar diffusion. The differential behavior between ions and electrons produces Hall effect. Partial-pressure gradients lead to a battery effect, etc. The importance of these non-ideal effects strongly depends on height and on the strength and gradients of solar magnetic fields. 

It can be expected from general considerations that non-ideal effects are less important in the photosphere, except for regions with strong concentrated magnetic field, such as sunspots and magnetic elements of the quiet regions \cite{Khomenko+etal2014b}. However, in the chromosphere, non-ideal effects may play a crucial role for its energy balance and formation of dynamic structures \cite{Krasnoselskikh2010, Khomenko+Collados2012, MartinezSykora+etal2012, Leake+Liton2013, Zaqarashvili+etal2013}. While it has long been known that a local thermodynamic equilibrium approach can not be applied to treat the processes related to radiative transfer in the chromosphere, it is now become evident that for the description of processes related to the interaction of chromospheric plasma with magnetic field, the standard magnetohydrodynamic approach may fail, and a new alternative description should be applied. A plausible alternative to the more complex kinetic approach is the application of multi-fluid theory where the individual species are described as fluids interacting by collisions. Depending on the degree of collisional coupling, it may be sufficient to apply a quasi-MHD single fluid theory complemented with the generalized Ohm's law, or to treat neutral and charged fluids separately. A number of works recently revise the multi-fluid formalism for the description of solar plasma  \cite{Meier+Shumlak2012,  Leake+etal2014, Khomenko+etal2014b}. Below we provide an overview of the recent advances in the field of multi-fluid modeling of the solar atmosphere. Previous reviews of the subject can be found in \cite{MartinezSykora2015, Khomenko2015, Leake2015}. 

\section{Wave propagation}
\label{sect:waves}

The effect of partial ionization on the propagation, excitation and damping of different types of waves in solar plasma has been addressed in many studies \cite{Piddington1956, Osterbrock1961, Haerendel1992, DePontieu1998, DePontieu+etal2001, Kumar+Roberts2003, Khodachenko2004, Khodachenko2006, Forteza2007, Pandey2008, Vranjes2008, Soler2009, Soler2010, Soler2013, Song+Vasyliunas2011,  Zaqarashvili2011, Zaqarashvili+etal2013}. In those works mostly an analytical approach has been applied, using either single or multiple fluid treatment. The effects of partial ionization have been found significant for high-frequency waves with frequencies close to the ion-neutral collision frequency, affecting the following aspects:
\begin{itemize}

\item Partial ionization produces additional {\it wave damping due to viscosity, conductivity, and friction} due to neutral component. The frictional damping is found to be more important for waves in the photosphere, chromosphere and prominence plasma \cite{Khodachenko2004, Khodachenko2006, Forteza2007}. The damping of waves depends both on the collisional frequency, and on the ionization fraction \cite{Kumar+Roberts2003, Pandey2008}. Ion-neutral friction provides an efficient way of dissipation of Alfv\'en and fast MHD waves \cite{Piddington1956, Osterbrock1961,  Haerendel1992,  DePontieu+etal2001, Leake+etal2005, Song+Vasyliunas2011, Soler2013, Soler+etal2014, Soler2015} that are significantly more difficult to dissipate compared to the compressional acoustic waves. This friction was proposed to be one of the reasons for the damping of coronal loop oscillations and prominence oscillations. On contrary, Hall effect affects waves mainly in the photosphere, where it can produce whistler waves and lead to instabilities \cite{Pandey2008, Pandey+Wardle2008}. 

\item It was suggested in \citep{Vranjes2008} that the {\it excitation rates of Alfv\'en waves} in the photosphere might be significantly decreased if the neutral component is taken into account. In the photosphere, the ion-neutral collisional frequency is expected to be larger than the cyclotron frequency and, therefore, perturbations due to the magnetic Lorentz force should also involve neutrals dragged by collisions with ions. It has been suggested that the perturbed velocity may be reduced, leading to a smaller Alfv\'en energy, since this flux is scaled with the factor $\rho_i/\rho_n$, of the ratio between the ion and neutral mass densities. However, a similar analysis by \cite{Tsap2011} contradicts this conclusion and presents an expression of the energy flux without the factor $\rho_i/\rho_n$.  Later on, \cite{Soler2013} demonstrated that the difference between both studies can be understood if one assumes that the initial perturbation involves neutrals as well, i.e. $v_n\ne0$, unlike \citep{Vranjes2008}.

\item Under a single fluid treatment, \cite{Soler2009, Barcelo2011} proposed that the collisional interaction with neutrals can lead to the {\it appearance of cut-off frequencies and wave numbers} for Alfv\'en and kink waves. However, as demonstrated by  \cite{Zaqarashvili2011}, the appearance of a cut-off wave number in the single-fluid approach might be due to the neglection of the inertia term in the equation of motion for the relative ion-neutral velocity. When this term is taken into account under a two-fluid treatment, no cut-off wave number owing to ion-neutral interaction is present. 

\item Apart from spatial and temporal damping effects (similar to the evanescence of acoustic-gravity waves in a stratified atmosphere), ion-neutral interaction can lead to {\it dissipation of perpendicular currents produced by waves}, converting the magnetic energy of waves into thermal energy and producing an important heating of the magnetized chromosphere above magnetic elements \cite{Goodman2000, Goodman2004, Goodman2010, Song+Vasyliunas2011}. This issue will be addressed in more detail in Section \ref{sect:heating}.

\end{itemize}

\section{Instabilities}

Ion-neutral interaction leads to a modification of the onset criteria and the growth rates of classical instabilities such as Rayleigh-Taylor (RTI) and Kelvin-Helmholtz (KHI), and also to the appearance of new ones, typical for a partially ionized medium: 

\begin{itemize}

\item {\it Hall instability in the presence of a flow shear}. This non-ideal magnetohydrodynamic instability develops due to Hall currents, produced because ions are collisionally dragged by neutrals, while electrons are bound to the magnetic field \cite{Vranjes+etal2006, Pandey+Wardle2012, Pandey+Wardle2013}. In the presence of a flow shear, this effect twists the radial magnetic field and generates an azimuthal field, while torsional oscillations of the azimuthal field generate again the radial field.  The maximum growth rate of the instability is proportional to the absolute value of the velocity shear and to ambient diffusivities, and occurs when both the field and the wave vector are vertical \cite{Pandey+Wardle2012, Pandey+Wardle2013}. When the field has a non-vertical component and when waves are propagating obliquely, both Hall and ambipolar diffusion together help the development of the instability.

\item {\it Farley-Buneman instability}. This two-stream instability arises under chromospheric conditions due to the drift motions of charged particles when electrons are strongly magnetized but ions are unmagnetized due to collisions with neutrals. It can be produced by waves or flows of quasi-neutral gas from the photosphere creating cross-field motion in a partially ionized plasma \cite{Fontenla2005, Fontenla2008, Gogoberidze+etal2009, Gogoberidze+etal2014, Madsen+etal2014}. Conditions in the chromosphere meet the instability criteria if the electron drift trigger velocity is slightly below the sound speed \cite{Madsen+etal2014}. It has been suggested that this instability may lead to chromospheric heating, however,  currents necessary to provide electron drift velocity of 2-4 km s$^{-1}$ are two orders of magnitude larger that actually measured \cite{Socas-Navarro2007, Gogoberidze+etal2009, Gogoberidze+etal2014}.

\item {\it Contact instabilities.} Magnetic RTI and KHI contact instabilities are expected to arise at the interfaces between the prominence and coronal material at the prominence boundary and in prominence threads. Linear analyses and numerical simulations  show that the presence of neutrals in a partially ionized plasma removes the critical wavelength imposed by the magnetic field, making perturbations unstable in the whole wavelength range \cite{Soler+etal2012, Diaz+etal2012, Diaz+etal2013, Khomenko+etal2014}, but with a small growth rate that depends on the ionization fraction. The instability threshold of the compressional KHI is very much sensitive to the value of the flow, and is lower than in the fully ionized case due to ion-neutral coupling, reaching sub-Alfv\'enic values \cite{Soler+etal2012}. Another kind of contact instability called dissipative instability was recently investigated in \cite{Ballai+etal2015} concluding that while viscosity tends to destabilise the plasma, the effect of partial ionization acts towards stabilizing the interface.

\end{itemize}

\section{Flux emergence}

Another phenomenon affected by ion-neutral interaction is the magnetic flux emergence \citep{Leake+Arber2006, Arber2007}. A magnetic flux rope rising due to buoyancy through the solar interior to the surface encounters a layer of almost neutral gas in the photosphere where it becomes affected by ambipolar diffusion removing perpendicular currents and modifying the structure of the emerged field. By performing 2D and 3D numerical simulations, \citep{Leake+Arber2006, Arber2007} have shown that the amount of the emerged flux can be greatly increased by the presence of this diffusive layer of partially ionized plasma. At the chromosphere, the order of magnitude stronger dissipation of currents perpendicular to the magnetic field, compared to that of longitudinal currents, has been found to facilitate the creation of potential force-free field structures  \cite{Arber2009}. Curiously, in the simulations by \citep{Leake+Arber2006, Arber2007} it was found that including neutrals avoids the Rayleigh-Taylor instability at the interface of the emerged flux. This apparently contradicts the results of linear analyses and numerical simulations of contact instabilities above \cite{Diaz+etal2013, Khomenko+etal2014}. This contradiction can be understood taking into account that, while the cut-off wavelength is removed in the partially ionized case, the small scale unstable modes are possibly not resolved in the simulations by \citep{Leake+Arber2006, Arber2007}  with much coarser numerical resolution compared to \cite{Khomenko+etal2014}.

\section{Magnetic reconnection}

The plasma partial ionization is also important for magnetic reconnection. While ambipolar diffusion affects only perpendicular currents, and therefore cannot directly produce reconnection, it can modify the configuration of the magnetic field lines, making the conditions for reconnection more favorable. Initial studies have shown that the reconnection rates indeed strongly depend on the collisional coupling between the ionized and neutral species  \cite{Zweibel1989}. Due to the action of ambipolar diffusion, oppositely oriented magnetic field lines can be brought sufficiently close to facilitate the reconnection  \cite{Brandenburg+Zweibel1994, Brandenburg+Zweibel1995}. The current sheet can become thin to the scale of the neutral-ion mean free path  \cite{Murphy+Lukin2015}. Two-fluid numerical model of reconnection applied in \cite{Sakai2006, Smith+Sakai2008, Sakai+Smith2009} demonstrated that reconnection leads to proton heating and jet-like phenomena with different temperatures of neutral and ionized species. A more advanced numerical modeling of reconnection in a Harris current including ion-neutral scattering collisions, ionization, recombination, optically thin radiative loss, collisional heating, and thermal conduction \cite{Leake+etal2012}  has shown a more complex picture where neutral and ionized fluid becomes uncoupled upstream from the reconnection site. It undergoes ion recombination that, combined with Alfv\'enic outflows, leads to fast reconnection rates independent of Lundquist number. Asymmetric reconnection is also possible to occur in the chromosphere when the emerged magnetic flux interacts with an already existing one. This situation was studied in \cite{Murphy+Lukin2015}, showing that the ion and neutral outflows are strongly coupled, but the inflows are asymmetrically decoupled. Yet another kind of reconnection induced by slow mode shock waves was modeled in \cite{Hillier+etal2016} who showed that the system undergoes several stages from weak to strong coupling finally reaching the quasi-steady state. It is characterized initially by an over-pressured neutral region that explosively expands outwards from the reconnection site, and frictional heating is produced across the shock front due to strong drift velocity. Altogether these models suggest that such kind of phenomena can be responsible for foot point heating of coronal loops, explosive events, chromospheric, transition region and sunspot penumbrae jets, and to produce spicules.

\section{Chromospheric heating}
\label{sect:heating}

From the point of view of the energy balance, additional dissipation of perpendicular currents due to ion-neutral interaction (ambipolar diffusion) can lead to a several orders of magnitude larger Joule heating compared to the fully ionized plasma \cite{Khodachenko2004, Arber2007}. It has been pointed out \cite{DePontieu1998, Judge2008, Krasnoselskikh2010} that current dissipation, enhanced by the presence of neutrals in a plasma not entirely coupled by collisions, can play an important role in the energy balance of the chromosphere and above. Numerical modeling has confirmed that, indeed, the amount of heating is sufficiently large and the time scales associated to it are sufficiently short so that it can easily compensate the radiative energy losses of the magnetized chromosphere and explain the chromospheric temperature increase \cite{Khomenko+Collados2012, MartinezSykora+etal2012}. The necessary condition for the heating is the existence of non force-free magnetic field, producing perpendicular currents.

Such currents can exist in the form of relatively stationary currents in magnetic structures, but can also be dynamically created by waves and flows. The magnetized plasma motions in the solar photosphere create an electromagnetic Poynting energy flux sufficient to heat the upper solar atmosphere \cite{Osterbrock1961,  Steiner2008, Goodman2010, Krasnoselskikh2010, Goodman2011, Abbett2012, Shelyag+etal2012}, and  a significant part of this flux may be produced in the form of Alfv{\'e}n waves \cite{Shelyag+Przybylski2014} (but see Section \ref{sect:waves}). A recent analysis by \cite{Shelyag+etal2016} has shown that up to 80\% of the Poynting flux associated to these waves can be dissipated and converted into heat due to the effect of ambipolar diffusion providing an order of magnitude larger amount of energy to the chromosphere compared to the dissipation of stationary currents \cite{Khomenko+Collados2012}. Nevertheless, \cite{Arber2016} argue that heating produced by acoustic shocks is more important than than by Alfv\'en wave dissipation through ion-neutral collisions. Therefore, the question of chromospheric heating due to the ion-neutral interaction will require further studies in the future.

\section{Concluding remarks}

Ion-neutral interaction in the partially ionized solar plasma can have significant effects on the dynamical processes and on the energy balance. In recent years, a mathematical formalism for the treatment of such interaction is being settled and analytical and numerical models of increasing complexity demonstrate how neutrals can affect wave propagation, plasma instabilities, reconnection and other fundamental processes taking place in the solar atmosphere. These effects are expected to influence the formation of chromospheric dynamic features, to affect its energy balance, the stability of prominences, etc. There is a need to step forward from single-fluid to multi-fluid modeling to allow relaxing the restrictions of such approximation and studying the uncoupled behavior of the ionized and neutral components of the solar plasma.

\section{Acknowledgements}
This work was supported by the Spanish Ministry of Science through the project AYA2014-55078-P and by the European Research Council in the frame of the FP7 Specific Program IDEAS through the Starting Grant ERC-2011-StG 277829-SPIA.
%

\providecommand{\noopsort}[1]{}\providecommand{\singleletter}[1]{#1}%
\providecommand{\newblock}{}


\begin{thebibliography}{10}
\expandafter\ifx\csname url\endcsname\relax
  \def\url#1{{\tt #1}}\fi
\expandafter\ifx\csname urlprefix\endcsname\relax\def\urlprefix{URL }\fi
\providecommand{\eprint}[2][]{\url{#2}}

\bibitem{Asplund+etal2000}
Asplund M, Nordlund {\AA}, Trampedach R, \mbox {Allende Prieto} C and Stein R~F
  2000 {\em A\&A\/} {\bf 359} 729---742

\bibitem{Khomenko+Collados2006}
Khomenko E and Collados M 2006 {\em ApJ\/} {\bf 653} 739---755

\bibitem{Cheung+etal2007}
{Cheung} M~C~M, {Sch{\"u}ssler} M and {Moreno-Insertis} F 2007 {\em A\&A\/}
  {\bf 467} 703--719

\bibitem{Moreno-Insertis+etal2008}
{Moreno-Insertis} F, {Galsgaard} K and {Ugarte-Urra} I 2008 {\em ApJ\/} {\bf
  673} L211--L214

\bibitem{Khomenko+etal2014b}
{Khomenko} E, {Collados} M, {D{\'{\i}}az} A and {Vitas} N 2014 {\em Physics of
  Plasmas\/} {\bf 21} 092901 

\bibitem{Krasnoselskikh2010}
{Krasnoselskikh} V, {Vekstein} G, {Hudson} H~S, {Bale} S~D and {Abbett} W~P
  2010 {\em ApJ\/} {\bf 724} 1542--1550 

\bibitem{Khomenko+Collados2012}
{Khomenko} E and {Collados} M 2012 {\em ApJ\/} {\bf 747} 87 

\bibitem{MartinezSykora+etal2012}
{Mart{\'{\i}}nez-Sykora} J, {De Pontieu} B and {Hansteen} V 2012 {\em ApJ\/}
  {\bf 753} 161

\bibitem{Leake+Liton2013}
{Leake} J~E and {Linton} M~G 2013 {\em ApJ\/} {\bf 764} 54 

\bibitem{Zaqarashvili+etal2013}
{Zaqarashvili} T~V, {Khodachenko} M~L and {Soler} R 2013 {\em A\&A\/} {\bf 549}
  A113 

\bibitem{Meier+Shumlak2012}
{Meier} E~T and {Shumlak} U 2012 {\em Physics of Plasmas\/} {\bf 19} 072508

\bibitem{Leake+etal2014}
{Leake} J~E, {DeVore} C~R, {Thayer} J~P, et al. 2014 {\em
  Space Sci.\ Rev.\/} {\bf 184} 107--172 

\bibitem{MartinezSykora2015}
{Mart{\'{\i}}nez-Sykora} J, {De Pontieu} B, {Hansteen} V and {Carlsson} M 2015
  {\em Philosophical Transactions of the Royal Society of London Series A\/}
  {\bf 373} 20140268--20140268 
  
\bibitem{Khomenko2015}
{Khomenko} E 2015  {\em Highlights of Spanish Astrophysics VIII\/} ed
  {Cenarro} A~J, {Figueras} F, {Hern{\'a}ndez-Monteagudo} C, {Trujillo Bueno} J
  and {Valdivielso} L pp 677--688 
  
\bibitem{Leake2015}
{Leake} J 2015 {\em IAU General Assembly\/} {\bf 22} 2289623

\bibitem{Piddington1956}
{Piddington} J~H 1956 {\em MNRAS\/} {\bf 116} 314

\bibitem{Osterbrock1961}
{Osterbrock} D~E 1961 {\em ApJ\/} {\bf 134} 347

\bibitem{Haerendel1992}
{Haerendel} G 1992 {\em Nat\/} {\bf 360} 241--243

\bibitem{DePontieu1998}
{De Pontieu} B and Haerendel G 1998 {\em ApJ\/} {\bf 338} 729

\bibitem{DePontieu+etal2001}
{De Pontieu} B, {Martens} P~C~H and {Hudson} H~S 2001 {\em ApJ\/} {\bf 558}
  859--871

\bibitem{Kumar+Roberts2003}
Kumar N and Roberts B 2003 {\em Solar Phys.\/} {\bf 214} 241

\bibitem{Khodachenko2004}
Khodachenko M~L, Arber T~D, Rucker H~O and Hanslmeier A 2004 {\em A\&A\/} {\bf
  422} 1073

\bibitem{Khodachenko2006}
Khodachenko M, Rucker H, Oliver R, Arber T and Hanslmeier A 2006 {\em Advances
  in Space Research\/} {\bf 37} 447

\bibitem{Forteza2007}
Forteza P, Oliver R and Ballester J~L 2007 {\em A\&A\/} {\bf 461} 731

\bibitem{Pandey2008}
Pandey B~P, Vranjes J and Krishan V 2008 {\em MNRAS\/} {\bf 386} 1635

\bibitem{Vranjes2008}
Vranjes J, Poedts S, Pandey B~P and Pontieu B~D 2008 {\em A\&A\/} {\bf 478} 553

\bibitem{Soler2009}
{Soler} R, {Oliver} R and {Ballester} J~L 2009 {\em ApJ\/} {\bf 699} 1553--1562
  
\bibitem{Soler2010}
{Soler} R, {Oliver} R and {Ballester} J~L 2010 {\em A\&A\/} {\bf 512} A28+

\bibitem{Soler2013}
{Soler} R, {Carbonell} M, {Ballester} J~L and {Terradas} J 2013 {\em ApJ\/}
  {\bf 767} 171 

\bibitem{Song+Vasyliunas2011}
{Song} P and {Vasyli{\= u}nas} V~M 2011 {\em Journal of Geophysical Research
  (Space Physics)\/} {\bf 116} A09104

\bibitem{Zaqarashvili2011}
{Zaqarashvili} T~V, {Khodachenko} M~L and {Rucker} H~O 2011 {\em A\&A\/} {\bf
  529} A82+ 
  
\bibitem{Leake+etal2005}
{Leake} J~E, {Arber} T~D and {Khodachenko} M~L 2005 {\em A\&A\/} {\bf 442}
  1091--1098 

\bibitem{Soler+etal2014}
{Soler} R, {Goossens} M, {Terradas} J and {Oliver} R 2014 {\em ApJ\/} {\bf 781}
  111 
  
\bibitem{Soler2015}
{Soler} R, {Ballester} J~L and {Zaqarashvili} T~V 2015 {\em A\&A\/} {\bf 573}
  A79 
  
\bibitem{Pandey+Wardle2008}
Pandey B~P and Wardle M 2008 {\em MNRAS\/} {\bf 385} 2269--2278

\bibitem{Tsap2011}
{Tsap} Y~T, {Stepanov} A~V and {Kopylova} Y~G 2011 {\em Solar Phys.\/} {\bf
  270} 205--211

\bibitem{Barcelo2011}
{Barcel{\'o}} S, {Carbonell} M and {Ballester} J~L 2011 {\em A\&A\/} {\bf 525}
  A60

\bibitem{Goodman2000}
{Goodman} M~L 2000 {\em ApJ\/} {\bf 533} 501--522

\bibitem{Goodman2004}
{Goodman} M~L 2004 {\em A\&A\/} {\bf 416} 1159--1178

\bibitem{Goodman2010}
{Goodman} M~L and {Kazeminezhad} F 2010 {\em ApJ\/} {\bf 708} 268--287

\bibitem{Vranjes+etal2006}
{Vranjes} J, {Pandey} B~P and {Poedts} S 2006 {\em Planetary and Space
  Science\/} {\bf 54} 695--700

\bibitem{Pandey+Wardle2012}
{Pandey} B~P and {Wardle} M 2012 {\em MNRAS\/} {\bf 426} 1436--1443

\bibitem{Pandey+Wardle2013}
{Pandey} B~P and {Wardle} M 2013 {\em MNRAS\/} {\bf 431} 570--581
  
\bibitem{Fontenla2005}
{Fontenla} J~M 2005 {\em A\&A\/} {\bf 442} 1099--1103

\bibitem{Fontenla2008}
{Fontenla} J~M, {Peterson} W~K and {Harder} J 2008 {\em A\&A\/} {\bf 480}
  839--846

\bibitem{Gogoberidze+etal2009}
{Gogoberidze} G, {Voitenko} Y, {Poedts} S and {Goossens} M 2009 {\em ApJ\/}
  {\bf 706} L12--L16 

\bibitem{Gogoberidze+etal2014}
{Gogoberidze} G, {Voitenko} Y, {Poedts} S and {De Keyser} J 2014 {\em MNRAS\/}
  {\bf 438} 3568--3576 

\bibitem{Madsen+etal2014}
{Madsen} C~A, {Dimant} Y~S, {Oppenheim} M~M and {Fontenla} J~M 2014 {\em ApJ\/}
  {\bf 783} 128

\bibitem{Socas-Navarro2007}
{Socas-Navarro} H 2007 {\em ApJS\/} {\bf 169} 439--457

\bibitem{Soler+etal2012}
{Soler} R, {D{\'{\i}}az} A~J, {Ballester} J~L and {Goossens} M 2012 {\em ApJ\/}
  {\bf 749} 163 

\bibitem{Diaz+etal2012}
{D{\'{\i}}az} A~J, {Soler} R and {Ballester} J~L 2012 {\em ApJ\/} {\bf 754} 41

\bibitem{Diaz+etal2013}
{D{\'{\i}}az} A~J, {Khomenko} E and {Collados} M 2014 {\em A\&A\/} {\bf 564}
  A97 
  
\bibitem{Khomenko+etal2014}
{Khomenko} E, {D{\'{\i}}az} A, {de Vicente} A, {Collados} M and {Luna} M 2014
  {\em A\&A\/} {\bf 565} A45 

\bibitem{Ballai+etal2015}
{Ballai} I, {Oliver} R and {Alexandrou} M 2015 {\em A\&A\/} {\bf 577} A82
  
\bibitem{Leake+Arber2006}
{Leake} J~E and {Arber} T~D 2006 {\em A\&A\/} {\bf 450} 805--818

\bibitem{Arber2007}
Arber T~D, Haynes M and Leake J~E 2007 {\em ApJ\/} {\bf 666} 541

\bibitem{Arber2009}
Arber T~D, Botha G~J~J and Brady C~S 2009 {\em ApJ\/} {\bf 705} 1183

\bibitem{Zweibel1989}
Zweibel E~H 1989 {\em ApJ\/} {\bf 340} 550

\bibitem{Brandenburg+Zweibel1994}
Brandenburg A and Zweibel E~H 1994 {\em ApJ\/} {\bf 427} L91

\bibitem{Brandenburg+Zweibel1995}
Brandenburg A and Zweibel E~H 1995 {\em ApJ\/} {\bf 448} 734

\bibitem{Murphy+Lukin2015}
{Murphy} N~A and {Lukin} V~S 2015 {\em ApJ\/} {\bf 805} 134 

\bibitem{Sakai2006}
Sakai J~I, Tsuchimoto K and Sokolov I~V 2006 {\em ApJ\/} {\bf 642} 1236

\bibitem{Smith+Sakai2008}
Smith P~D and Sakai J~I 2008 {\em A\&A\/} {\bf 486} 569

\bibitem{Sakai+Smith2009}
Sakai J~I and Smith P~D 2009 {\em ApJ\/} {\bf 691} L45

\bibitem{Leake+etal2012}
{Leake} J~E, {Lukin} V~S, {Linton} M~G and {Meier} E~T 2012 {\em ApJ\/} {\bf
  760} 109

\bibitem{Hillier+etal2016}
{Hillier} A, {Takasao} S and {Nakamura} N 2016 {\em A\&A\/} {\bf 591} A112
 
\bibitem{Judge2008}
{Judge} P 2008 {\em ApJ\/} {\bf 683} L87--L90 

\bibitem{Steiner2008}
{Steiner} O, {Rezaei} R, {Schaffenberger} W and {Wedemeyer-B{\"o}hm} S 2008
  {\em ApJ\/} {\bf 680} L85 

\bibitem{Goodman2011}
{Goodman} M~L 2011 {\em ApJ\/} {\bf 735} 45 

\bibitem{Abbett2012}
{Abbett} W~P and {Fisher} G~H 2012 {\em Solar Phys.\/} {\bf 277} 3--20
  
\bibitem{Shelyag+etal2012}
{Shelyag} S, {Mathioudakis} M and {Keenan} F~P 2012 {\em ApJ\/} {\bf 753} L22

\bibitem{Shelyag+Przybylski2014}
{Shelyag} S and {Przybylski} D 2014 {\em Publications of the Astronomical
  Society of Japan\/} {\bf 66} S9 

\bibitem{Shelyag+etal2016}
{Shelyag} S, {Khomenko} E, {de Vicente} A and {Przybylski} D 2016 {\em ApJ\/}
  {\bf 819} L11 

\bibitem{Arber2016}
{Arber} T~D, {Brady} C~S and {Shelyag} S 2016 {\em ApJ\/} {\bf 817} 94

\end{thebibliography}
\end{document}